\documentstyle[12pt]{article}

\begin{document}

\thispagestyle{empty} %{\hfill  Preprint JINR E2-93-358}\vspace{2.5cm} \\
\vspace*{3cm}

\begin{center}
{\Large {\bf From Antibracket to Equivariant Characteristic Classes }}\\

\bigskip
\vspace*{2.4cm}
{\large {\bf Armen Nersessian}}
\footnote{e-mail:nersess@theor.jinrc.dubna.su
nerses@thsun1.jinr.dubna.su}%
\footnote{Supported in part by Grant No.
M21000 from International Science Foundation} \vspace*{0.8cm}\\ {\it
Bogoliubov Theoretical Laboratory , JINR}\\ {\it %
Dubna, Moscow Region 141980, Russia}\\
\end{center}
  \bigskip \bigskip \bigskip \bigskip \bigskip

\begin{abstract}
We construct  the
odd symplectic structure  and  the equivariant even (pre)symp\-lec\-tic
one from it on the space of differential forms on the Riemann manifold.
The Poincare -- Cartan like invariants of the second structure
 define the equivariant
generalizations of the Euler classes on the surfaces.
\end{abstract}

\bigskip
\bigskip
\bigskip
\bigskip

\setcounter{page}0
\newpage
\setcounter{equation}0

\section{Introduction}

Equivariant cohomology (see e. g. \cite{DH}-\cite{berline}) is presently
attracting much interest in physics. It is stimulated by the applications of
localization formulas to evaluation of path integrals
for a wide class of field-theoretical systems
(see e. g. \cite{niemi}, \cite{niemi2} and refs. therein) including
 {\it e. g.} topological field theories \cite
{aj,witten,blau},  and supersymmetric \cite{morozov} theories.

Equivariant cohomology nicely formulated by using the language of
supermathematics, where the role of supermanifolds played by the
space of differential forms $\Lambda(M)$ on the given manifold $M$
(correspondingly, the role of the odd variables on $\Lambda(M)$
is played by the basic 1-forms).

For the equivariant localization of the integrals over sub-manifolds
$N\subset M$, it is nesessary to construct the equivariant generalization
 of the characteristic classes on $N$.
Such a generalizations of the Eiler and Cartan classes are known
(see , e. g., \cite{berline,aj}. From supergeometrical point of view,
thats define the integral densities  on the space of differential forms
$\Lambda(N)\subset\Lambda(M)$ on $N$.

However, in supersymmetric theories the odd coordinates play the roly of
dynamical, but not auxillary variables. So, the sub-(super)space
$\Gamma\subset\Lambda(M)$ of
the physical fields can be have the structure of supermanifold, differed
from $\Lambda(N)$.

In this paper we construct a family of the equivariant integral densities
on the such a surfaces $\Gamma$, which generalize the known construction
of the equivariant Euler classes.

For this purpose we firstly construct the odd symplectic structure $\Omega_1$
on the space of differential forms $\Lambda (M)$ on the Riemann manifold $M$ .
The Lie derivative of $\Omega_1$ along the vector field,
corresponding to $S^1$- equivariant transformation (
where $S^1-$ action on $M$  defines the isometry of metrics )
is the {\it $S^1$-equivariant even
(pre)symplectic structure} $\Omega_0$ ({\it Section 2}).

The Poincare- Cartan like
invariants \cite{ks} of $\Omega_0$ define the set of equivariant
densities, generalizing the known  equivariant characteristic classes
 \cite{berline}
({\it Section 3}).

Notice, that initial object in our considerations : the odd symplectic
structure is used mainly in the Batalin-Vilkovisky  quantization scheme
\cite{bat}. However, the recent investigations of its geometry
\cite{BVgeom} allow to assume one that the odd symplectic structure plays the
essential role in
problems connected with the integration over (super)surfaces.

Notice also that in \cite{jetp} it has been demonstrated that the
odd symplectic structure, constructed on the space of a differential form
on the symplectic manifold, naturally describes its  equivariant
cohomologies  and establishes the correspondence of the
equivariant cohomologies
to the bi-Hamiltonian supersymmetric dynamics (with even and odd symplectic
structures) \cite{asi} .
\setcounter{equation}0

\section{Odd and Even Symplectic Structures}

In this section we  construct   the odd symplectic structure  and then
the even $S^1$-invariant (pre)symplectic one on the space of
differential forms on the
Riemann manifold $M$ .

Let $(M, g )$ be the Riemann manifold  and $\xi$  its Killing vector
defining the $S^1$ action.
Let $\Lambda (M)$ be the space of differential
forms on $M$.
It can be parametrized by the local coordinates $z^A =(x^i
, \theta^i)$ , where $x^i$ denote the local coordinates on $M$;
 and $\theta^i$ denote the basic 1-forms $dx^i$, $p(\theta^i)= 1 $ .

Consider the vector fields $\hat X$ and $\hat E$  on $\Lambda (M)$:
\begin{equation}
\label{E}{\hat X}=\xi^i\frac{\partial}{\partial x^i} + \xi_{,k}^{i}\theta^k
\frac{\partial}{\partial \theta^i},\quad {\hat E} =\xi^i \frac{\partial}{%
\partial \theta^i} +\theta^i \frac{\partial}{\partial x^i} :\quad  [\hat E ,
\hat E ]_{+} =2\hat X .
\end{equation}
It is obvious that $\hat X$ corresponds to the  Lie derivative
of differential forms on $M$ along $\xi$: $\hat X\rightarrow L_{\xi}
$, and $\hat E$ corresponds to the  $\xi$-equivariant ($S^1$-equivariant
) differentiation on $M$ :
 ${\hat E}\rightarrow d_\xi=d +\imath_\xi$ .
The last expression in (\ref{E}) corresponds to the
homotopy formula $ L_{\xi}= d\imath_{\xi}
+\imath_{\xi} d  $.\\
Below we  consider $\Lambda (M)$ as a supermanifold
 and denote by ${\cal L}$ and $d$
the Lie derivative and exterior differentiation
on $\Lambda (M)$ respectively  .\\
It is easy to see that the Berezin integration on $\Lambda(M)$
leads to the  integration of differential forms on $M$. \\

Let us construct on $\Lambda (M)$ the odd symplectic structure taking in the
coordinates $(x^i, \theta^i)$  the form
\begin{equation}
\label{ogs} \Omega_{1}= dx^i\wedge d(g_{ij}\theta^{j}) =g_{ij}dx^{i}\wedge
D\theta^{j} , \quad D\theta ^i=d\theta  ^i+\Gamma _{kl}^i\theta ^kdx^l,
\end{equation}
where $\Gamma _{kl}^i$ are the Cristoffel symbols for the metric $g_{ij}$.\\
The corresponding odd Poisson bracket (antibracket ) is :
\begin{equation}
\label{ob} \{f,g\}_1 =g^{ij} (\nabla_{i}f\frac{\partial g}{%
\partial\theta^j} - \frac{\partial f}{\partial\theta^i} \nabla_{j}g ) ,
\quad\nabla_{i}=\frac{\partial}{\partial x^{i}}- \Gamma _{ik}^j\theta ^k
\frac{\partial_{l}}{\partial \theta^{j}} ,
\end{equation}
It satisfies the conditions
\begin{eqnarray}
  & &\{ f, g \}_{1} =
-(-1)^{(p(f)+1)(p(g)+1)}\{ g, f \}_{1} \quad {\rm ( "antisymmetricity")}
,\label{anti}\nonumber \\
 & & (-1)^{(p(f)+1)(p(h)+1)}\{ f,\{ g, h \}_{1}\}_{1} +
{\rm {cycl. perm. (f, g, h)}} = 0 \quad{\rm {(Jacobi\quad id.)}} .
\label{bjac}\nonumber \end{eqnarray}
The odd symplectic structure (\ref{ogs}) is $\hat X$-invariant
, so $\hat X$
can be presented in the Hamiltonian form
\begin{equation}
\label{HL}\hat X = \{ Q_1, .\}_{1} ,\quad {\rm where }\quad
Q_1=\xi_{i}\theta^{i}.
\end{equation}
But it is not $\hat E$-invariant:
$$
{\cal L}_{E}\Omega_{1}={\tilde \Omega }_0 \neq0 .%
$$
Here
\begin{equation}
\label{evens}\Omega_0  = \frac 12 (\xi_{i;j}+g_{in}R^n_{jkl}\theta
^k\theta ^l)dx^i\wedge dx^j+g_{ij} D\theta^i\wedge D\theta ^j
\end{equation}
( $R^n_{jkl}$ is the curvature tensor on $M$ ), being $E$-invariant
 (i.e $S^1$- equivariant) even closed 2-form :
$$
p(\Omega_0) =0, \; d = {\cal L}_{E}d\Omega_{1}=0 ,\;\;\;\; {\cal L}%
_{E}\Omega_{0}= {\cal L}_{E}{\cal L}_{E}\Omega_{1}= 2{\cal L}_{X}\Omega_{1}
=0.%
$$
Therefore, the vector fields (\ref{E}) are Hamiltonian ones under
 $\Omega_0$. The corresponding
 Hamiltonians are
\begin{equation}
\label{Qxi}{\cal H}\equiv {\cal L}_{E} Q_1= \xi^i
g_{ij}\xi^{j}-\xi_{i;j}\theta^i\theta^j ,\quad {Q}_{2}=
\xi^{i}\xi_{i;j}\theta^j .
\end{equation}
The potential 1-form ${\cal A}$ : $d{\cal A} = \Omega_{0}$ is
\begin{equation}
\label{A} {\cal A}= \Omega_1 (\hat E , ...) =\xi_i dx^i +\theta^i
g_{ij}D\theta^j.
\end{equation}
{\it Thus, starting from the odd symplectic
structure  we
constructed the $S^1$-equivariant even (pre)symplectic structure
on the space of differential forms on the Riemann manifold.}
\setcounter{equation}0
\section{Equivariant Characteristic classes}
In this Section we  construct the equivariant characteristic
classes for the surfaces in $\Lambda (M)$.

Let $\Gamma\subset\Lambda (M) $ be a closed surface  and $\Omega_0$ be
nondegenerate on it.
Let $\Gamma$ is parametrized by the equations $%
z^{A}=z^{A}(w)$,  where $w^{\mu}$ are local coordinates of $\Gamma$.\\
Thus the following density
is correctly define on $\Gamma$
\begin{equation}
\label{d}{\cal D}_{\Gamma}(w)=\sqrt{{\rm Ber}\;\Omega_0\vert_{\Gamma}}
\equiv \sqrt{{\rm Ber}\frac{\partial _r z^A}{\partial w^{\mu}}
\Omega_{(0)AB}\frac{\partial _l z^B}{\partial w^{\nu}}},
\end{equation}
and it  is invariant under canonical transformations
of the presymplectic structure (\ref{evens})
 \cite{ks}. So, this density is  $S^1$-equivariant too. \\
Hence, {\it the
functional }
\begin{equation}
\label{invint} Z^{\lambda}(\Gamma, F )= \int_{\Gamma}{\rm e}%
^{F-\lambda{\hat E}\Psi} {\cal D}_{\Gamma}[dw],
\end{equation}
where $F(z)$ and $\Psi (z)$ are correspondingly the even $\hat E$- and odd
$\hat
X$-invariant functions
\begin{equation}
{\hat E} F =0, p(F) =0,\quad\ {\hat X}\Psi =0, p(\Psi)=1 ,
\end{equation}
{\it is $S^1$-equivariant for any compact $\Gamma$ }.
Therefore it is $\lambda$ -independent. \\

Let the 2-form (\ref{evens}) be nondegenerate  on
$\Lambda(M)$ and the surface $ \Gamma$ be defined by the equations $f^a
(z)=0$, $a=1,...{\rm codim}\Gamma $. In this case,  the functional
(\ref{invint}) can be  presented in the form (compare with \cite{km})
\begin{equation} \label{dinvint} Z^{\lambda}(\Gamma, F )=
\int_{\Lambda(M)}{\rm e}^{F(z)-\lambda ({\hat E}\Psi )} \delta
(f^a)\sqrt{{\rm Ber}\{f^a, f^b\}_0} {\cal D}_{0} dz ,\end{equation}
 where
\begin{equation}
\label{epb}\{f(z),g(z)\}_0 =\nabla _if(z)(\xi_{i;j} +R_{ijkl}\theta ^k\theta
^l)^{-1}\nabla _jg(z)+\frac 12\frac{\partial _rf(z)}{\partial \theta ^i}%
g^{ij}\frac{\partial _lg(z)}{\partial \theta ^j},
\end{equation}
is the Poisson bracket, corresponding to (\ref{evens}) and
${\cal D}_{0}(z)\equiv{\cal D}_{\Lambda(M)}(z) = \sqrt{
Ber\Omega_{(0)AB}}$ . \\
This functional is invariant both under
reparametrization of $\Lambda (M)$, and choice of the  functions
$f^a$. \\

The functional $Z^\lambda(\Gamma, 0)$ is invariant under smooth
deformations of $\Gamma$  ( if basic manifold $N$ of $\Gamma$
is closed )
\cite{km}, i. e., it is a
topological invariant of $\Gamma$ . \\
For example, $Z^\lambda(\Lambda(M), 0 )$ coincide with the Euler
number of $M$. In the limit $\lambda\to 0$ it gives the
Poincare-Hopf formula, and in the limit $\lambda\to \infty$ (where we
substitute $\Psi=Q_{1}=\xi_i\theta^i$),
Gauss--Bonnet one for the Euler number of $M$ \cite{niemi2}.\\
Hence, (\ref{d}) defines the $S^1$-equivariant characteristic
class of $\Gamma$.

{\bf Example :} Let $\Gamma\subset \Lambda(M)$ be associated with the vector
bundle $V(N)$ : $V(N)\subset T(M), N\subset M$. \\ Let it be parametrized by
the equations :
\begin{equation}
x^i=x^i (y^a), \quad \theta^i = P^i_\alpha(y)\eta^a,
\end{equation}
where $w^\mu =(y^a ,\eta^\alpha )$ are local coordinates of
$\Gamma$, $ p(y)=0 ,p(\eta)=1$
( $y^a$ are local coordinates of $N$) . \\
 Thus
\begin{equation}
\Omega_0\vert_{\Gamma} = \frac 12 (\xi_{[a,b]}+
g_{\alpha\delta}R^{\delta}_{\beta ab}\eta^{\alpha}\eta^{\beta} )dy^a\wedge
dy^b + g_{\alpha\beta}D\eta^{\alpha}\wedge D\eta^{\beta}.
\end{equation}
Here we introduce the notation :
$$
\xi_{[a,b]} =\xi_{i;j}\frac{\partial x^i}{\partial y^a} \frac{\partial x^j}{%
\partial y^b};\;\;\;g_{\alpha\beta}=P^i_\alpha g_{ij}P^j_\beta ;\;\;\;
D\eta^\alpha =d\eta^\alpha + A^\alpha_{a\beta}\eta^\beta dy^a ,%
$$
where
$$
A^\alpha_{a\beta} =g^{\alpha\delta}P^i_\delta g_{ij}\left
(P^j_{\beta,a}+\Gamma^j_{lk}P^k_\beta \frac{\partial x^l}{\partial y^a}%
\right)
$$
is the induced connection on $V(N)$ ( compatible with $g_{\alpha\beta}$ )
and $R^{\delta}_{\beta ab}$ is its curvature tensor. \\
Hence
\begin{equation}
\label{d1}{\cal D}_{\Gamma}(w)= \sqrt{{\rm Ber} \Omega_0
\vert_{\Gamma}} =\left(\frac{\det ( \xi_{[a,b]}+
g_{\alpha\delta}R^{\delta}_{\beta ab}\eta^{\alpha}\eta^{\beta})}{\det
g_{\alpha\beta}}\right)^\frac{1}{2}
\end{equation}
defines the family of equivariant characteristic classes of $N$.

In the case $\Gamma =\Lambda(N)$, i. e. $P^i_\alpha = \frac{\partial x^i
}{\partial y^a}$, (\ref{d1}) coincides with the known  equivariant
Euler classes on $N$  \cite{berline}.


\begin{thebibliography}{99}
\bibitem{DH}   Duistermaat J.J.,  Heckman G.J., {\it Inv. Math.}
{\bf 69}, 259 (1982)
 ;  ibid {\bf 72}, 153 (1983)

Atiah M. F.,  Bott R. {\it Topology}, {\bf 23}, 1 (1984)
\bibitem{mq}Matthai V., Quillen D., {\it Topology} {\bf 25}, 85 (1986),
\bibitem{berline}  Berline N., Getzler E., Vergne M., {\it Heat Kernel
and Dirac Operators}", Springer Verlag , Berlin, 1991
\bibitem{aj} Atiah M. F.,  Jefferey L., {\it J. Geom. Phys.}, {\bf 7},  119
 (1990)
\bibitem{witten}  Witten  E., {\it J. Geom. Phys.}, {\bf 9}, 303 (1992)
\bibitem{blau}  Blau M., {\it Matthai-Quillen formalism in Topological Field
Theories .}, hep-th 9203026
 \bibitem{morozov}  Morozov A.Yu., Niemi A. J., Palo
K. {\it Nucl.  Phys.}  {\bf B377 }, 295 (1992)
\bibitem{niemi} Niemi A. J.,  Tirkkonen O., {\it Phys. Lett.}
 {\bf B293}, 339 (1992); {\it Ann.  Phys.}  (N.  Y) (1994) (to appear)
\bibitem{niemi2}  A. J. Niemi, K.
Palo - Equivariant Morse Theory and Quantum Integrability. Preprint UU-ITP
10/94
 \bibitem{ks}
Khudaverdian O. M.,   Schwarz A.  S.,   Tyupkin Yu. S., {\it
Lett.  Math. Phys.}, {\bf 5}, 517 (1981)
\bibitem{bat}  Batalin I. A.,  Vilkovisky G. A., {\it Phys. Lett.},
{\bf B102}, 27 (1981); {\it Phys.  Rev} {\bf D28}, 2563 (1983) 2563
\bibitem{BVgeom}   Schwarz  A.,  {\it Comm. Math. Phys.} {\bf 155}, 249 (1993)
\bibitem{jetp}  Nersessian  A.,  {\it JETP Lett.}, {\bf 58}, 66  (1993)
\bibitem{asi}  Nersessian A., {\it Hamiltonian Mechanics: Integrability
and Chaotic Behavior} (ed. J. Seimenis),
 NATO ASI Series B:  Physics , Plenum Publ., 1994, p.353
\bibitem{km}   Khudaverdian O. M.,   Mkrtchian R.  L.,{\it Lett. Math.
Phys.}, {\bf  18}, 229 (1989)  \end{thebibliography}
 \end{document}